\documentclass[%
 reprint,
superscriptaddress,
 amsmath,amssymb,
 aps,
pra,
]{revtex4-2}
\usepackage[utf8]{inputenc}
\usepackage{amsmath, amssymb}
\usepackage{graphicx}
\usepackage{xcolor}
\usepackage{hyperref}
\usepackage{bm}
\usepackage{dcolumn}
\usepackage{subcaption}
\hypersetup{colorlinks=true, linkcolor=blue, citecolor=blue, urlcolor=blue}
\begin{document}

\title{Explicit Quantum Search Algorithm for the Densest $k$-Subgraph Problem}
\author{Yu.A. Biriukov}
\altaffiliation[e-mail: ]{biriukov.ia18@physics.msu.ru}
\affiliation{
 Quantum Technology Centre and Faculty of Physics, M.V. Lomonosov Moscow State University, 1 Leninskie Gory, Moscow 119991, Russia
}
\author{R.D. Morozov}
\thanks{These authors contributed equally to this work.}
\affiliation{
 Quantum Technology Centre and Faculty of Physics, M.V. Lomonosov Moscow State University, 1 Leninskie Gory, Moscow 119991, Russia
}

\author{I.V. Dyakonov}
\affiliation{
 Quantum Technology Centre and Faculty of Physics, M.V. Lomonosov Moscow State University, 1 Leninskie Gory, Moscow 119991, Russia
}
\affiliation{
Russian Quantum Center, 30 Bolshoy bul'var building 1, Moscow 121205, Russia
}

\author{S.S. Straupe}
\affiliation{
 Quantum Technology Centre and Faculty of Physics, M.V. Lomonosov Moscow State University, 1 Leninskie Gory, Moscow 119991, Russia
}
\affiliation{
Russian Quantum Center, 30 Bolshoy bul'var building 1, Moscow 121205, Russia
}

\begin{abstract}
This paper addresses the problem of finding the densest $k$-vertex subgraph in an arbitrary graph. This problem is NP-hard and has important applications in social network analysis, fraud detection, recommendation systems, and bioinformatics. We propose two quantum approaches to solve this problem: reduction to Quadratic Unconstrained Binary Optimization (QUBO) and using Grover's quantum search algorithm. For the latter approach, we present an explicit gate-based oracle circuit utilizing Dicke states and Quantum Fourier Transform for edge counting. Numerical simulations demonstrate a quadratic speedup over classical Brute-force search.
\end{abstract}

\maketitle

\section{Introduction and Problem Statement}

The Densest $k$-Subgraph (DkS) problem represents a fundamental challenge in graph theory and combinatorial optimization with significant practical applications across multiple domains. This problem involves finding a subset of $k$ vertices in an undirected graph $G = (V, E)$ that induces the maximum number of edges within the resulting subgraph. Formally, given a graph $G = (V, E)$ with $|V| = n$ vertices, we seek to identify a subset $S \subseteq V$ such that $|S| = k$ and the number of edges in the induced subgraph $G[S]$ is maximized. The adjacency matrix $A$ of the graph, where $A_{ij} = 1$ if edge $(i,j) \in E$ and $A_{ij} = 0$ otherwise, provides the complete structural information needed for this optimization.

The computational complexity of this problem places it in the NP-hard category, rendering exact solutions computationally intractable for large instances using classical algorithms. This inherent difficulty has motivated research into alternative computational paradigms, including quantum computing approaches that may offer advantages for such combinatorial optimization challenges. The practical significance of the Densest $k$-Subgraph problem extends to numerous real-world applications, including fraud detection in financial networks and online review systems \cite{Hooi2016, Li2020, Ji2022}, community identification in social networks \cite{Angel2014}, and functional module detection in biological networks \cite{Saha2010}. These diverse applications share the common need to identify densely connected substructures within larger networks, making efficient algorithms for this problem highly valuable.

Until recently, the primary approach for solving the Densest $k$-Subgraph problem on quantum computers has been through reformulation as a Quadratic Unconstrained Binary Optimization problem (QUBO) \cite{Calude2020, Huang2020}. This method maps the graph problem to an optimization format executable on quantum annealers such as D-Wave systems, leveraging the natural tendency of quantum systems to find low-energy states corresponding to optimal solutions. These QUBO-based formulations enable the direct use of annealing hardware \cite{pelofske2021decomposition} but also suffer from practical limitations, including dense connectivity requirements that lead to long transpilation circuits and degraded solution quality on current-generation devices \cite{calude2020quantum}.

Beyond annealing-based approaches, circuit-model quantum algorithms have been explored. In particular, Quantum Approximate Optimization Algorithm variants tailored to constrained optimization, including densest $k$-subgraph and related problems, have shown numerical evidence of favorable scaling of the required circuit depth, with logarithmic or low-degree polynomial dependence on $n$ compared to Grover-type unstructured search over the constrained subspace \cite{Golden2023}. A complementary line of work proposes learning a parameterized Grover oracle for constrained binary optimization, effectively training a variational quantum circuit so that a single Grover iteration prepares near-optimal solutions with high probability \cite{Ohno2024}. While these circuit-model approaches demonstrate promising behavior in simulation, they remain heuristic, and their complexity guaranties are either based primarily on numerical fitting.

In this work, we investigate alternative quantum approaches beyond QUBO reformulation, exploring whether other quantum algorithmic techniques can provide advantages for solving the Densest $k$-Subgraph problem. Specifically, we develop a novel quantum algorithm based on Grover's search framework \cite{grover1996fast} that provides a provable quadratic speedup over classical Brute-force search. The main advantage of our approach is an explicit low-depth gate scheme for the oracle, which is the main obstacle to the implementation of Grover's algorithm in practice. We use effective scheme for the preparation of the superposition of all subgraphs with k vertices, which turns out to be a Dicke state. Our oracle computes the number of edges using QFT-based arithmetic and marks the subgraphs that have more edges than a certain threshold. Finally, to finish an iteration of Grover's algorithm we need to perform the reflection in solutions space, which involves the construction of special diffusion operator. By gradually raising the threshold we reach the situation when only the subgraph with the largest number of edges is marked. This gives us the solution to the DkS problem. \\
The complete quantum circuit is explicitly constructed with detailed resource analysis, providing clear guidelines for implementation. Our research aims to broaden the toolkit available for solving combinatorial optimization problems on emerging quantum hardware by developing approaches that may offer different trade-offs in terms of circuit depth, number of qubit and computational efficiency. By examining methods that complement or potentially surpass existing Hamiltonian-based and variational approaches \cite{Calude2020, Huang2020, Golden2023, Ohno2024}, we contribute to the expanding landscape of quantum algorithms for graph optimization problems and help identify the most promising directions for future research and implementation.

\section{QUBO Method}

The QUBO problem has the form:
\begin{equation}
    \min_{x \in \{0,1\}^n} \sum_{i} Q_{ii} x_i + \sum_{i < j} Q_{ij} x_i x_j.
\end{equation}
To reduce the DkS problem to QUBO, we introduce binary variables $x_i$, where $x_i = 1$ indicates that vertex $i$ is included in the subset $S$. The objective function maximizing the number of edges is written as:
\begin{equation}
H = -\sum_{i < j} A_{ij} x_i x_j.
\end{equation}
The constraint $\sum_i x_i = k$ is incorporated using a penalty term:
\begin{equation}
\lambda \left( \sum_i x_i - k \right)^2.
\end{equation}
The complete objective function becomes:
\begin{equation}
H = -\sum_{i < j} A_{ij} x_i x_j + \lambda \left( \sum_i x_i - k \right)^2.
\end{equation}
Expanding the square and ignoring the constant $\lambda k^2$, we obtain:
\begin{equation}
H = \sum_i \left[ \lambda (1 - 2k) \right] x_i + \sum_{i < j} (2\lambda - A_{ij}) x_i x_j.
\end{equation}
The coefficient $\lambda$ must be sufficiently large to enforce the constraint. It is sufficient to choose $\lambda > \frac{k(k-1)}{2}$, as this is the maximum possible number of edges in a $k$-vertex subgraph. This approach was implemented on the D-Wave quantum annealer in \cite{Calude2020}.

\section{Quantum Search Algorithm}

The quantum search approach leverages a Grover-type search procedure to achieve quadratic speedup over Brute-force search. The algorithm starts from an initial threshold $m_0$, which serves as a lower-bound estimate for the number of edges in the optimal $k$-vertex subgraph. In this work, $m_0$ is chosen from the average edge density of the graph

\begin{equation}
    m_0 = \left\lfloor\frac{|E|}{\binom{n}{2}}\binom{k}{2}\right\rfloor.
\end{equation}
This value equals the expected number of edges in a random $k$-vertex induced subgraph and provides a simple starting point for the quantum search procedure.

\begin{figure}
    \centering
    \includegraphics[width=1\linewidth]{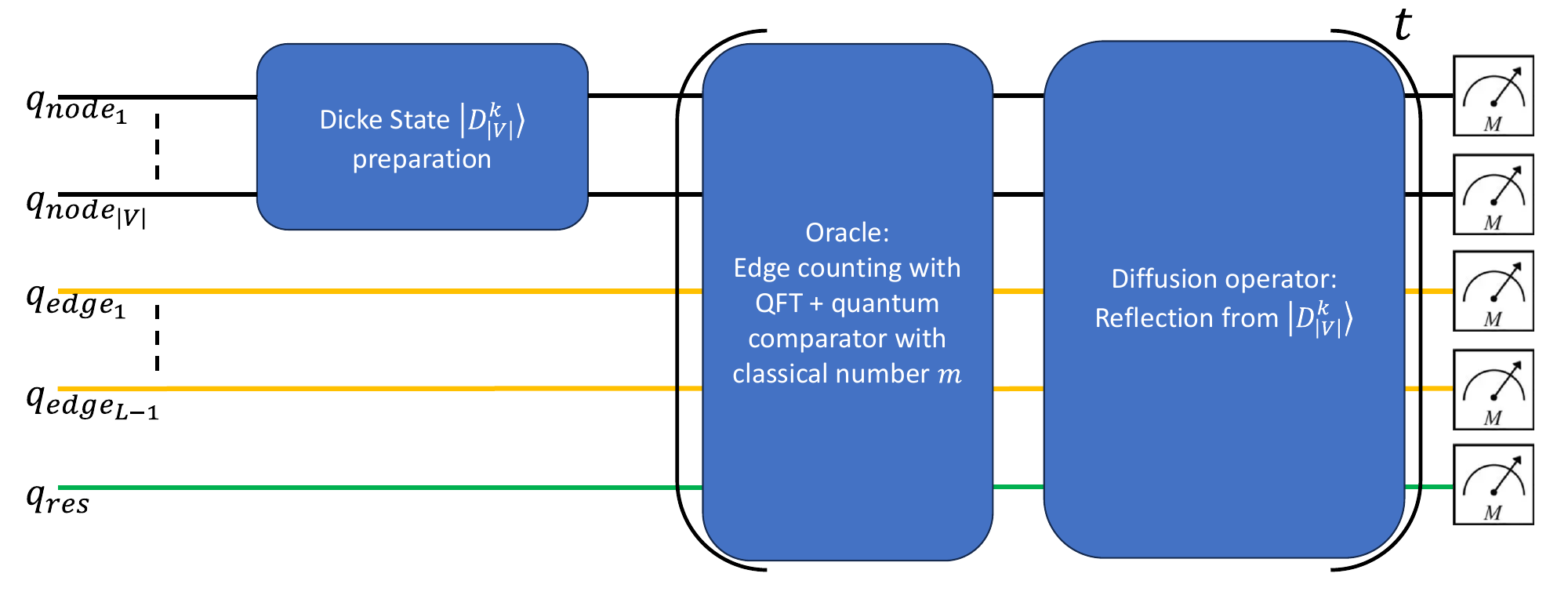}
    \caption{General scheme of a single Grover-based search call at threshold $m_i$.}
    \label{fig:gen_scheme}
\end{figure}

The core of the algorithm operates through an iterative search process over a sequence of threshold values $m_i$ that monotonically increase beyond the previous best solution. At iteration $i$ we fix a current threshold $m_i$ and run the Grover-based search with an oracle that marks $k$-vertex subgraphs whose edge count is at least $m_i$. After each Grover call the measured subgraph is checked classically: if its number of edges is strictly greater than $m_i$, we accept it as an improvement and update the threshold via $m_{i+1} = \mathrm{edges}(S)$, then restart the search at the new level. If the measured subgraph has at most $m_i$ edges, the attempt is counted as a failure. Once we observe $R$ consecutive failures at the same threshold level (with $R$ chosen as discussed below from the Dürr–Høyer analysis), we conclude with high confidence that no denser $k$-subgraph exists and terminate the algorithm. The general gate scheme of one iteration of search is depicted in Fig.~\ref{fig:gen_scheme}.

At a fixed threshold value $m_i$, a single Grover-based search call proceeds as follows. The circuit first prepares the Dicke state $|D^n_k\rangle$ in $q_{node}$ register, which represents all $k$-vertex subgraphs in superposition. A Dicke state is defined as the equal-amplitude superposition of all $n$-qubit computational basis states of Hamming weight $k$,
\begin{equation}
|D^n_k\rangle = \binom{n}{k}^{-\frac{1}{2}} \sum_{x \in \{0,1\}^n,\:\mathrm{HW}(x)=k} |x\rangle,
\end{equation}
where $\mathrm{HW}(x)$ denotes the Hamming weight of the bitstring $x$ (the number of ones in $x$). After this preparation, the circuit applies a prescribed number of Grover iterations combining the edge-counting oracle and the diffusion operator, and finally measures the vertex register to obtain a candidate $k$-subgraph for classical verification. In what follows we first describe the Dicke-state preparation routine and then turn to the oracle and diffusion operator.


Dicke states provide a uniform superposition over all $\binom{n}{k}$ $k$-vertex subsets. In our implementation we use the short-depth Dicke-state circuits of Bärtschi and Eidenbenz~\cite{Bartschi2022}, adapted to the all-to-all connectivity setting relevant for our problem. This construction prepares $|D^n_k\rangle$ with a total two-qubit gate count of $O(kn)$ and circuit depth $O\!\bigl(k \log (n/k)\bigr)$, which is sufficient for the overall algorithmic viability.

\begin{figure}
    \centering
    \includegraphics[width=1\linewidth]{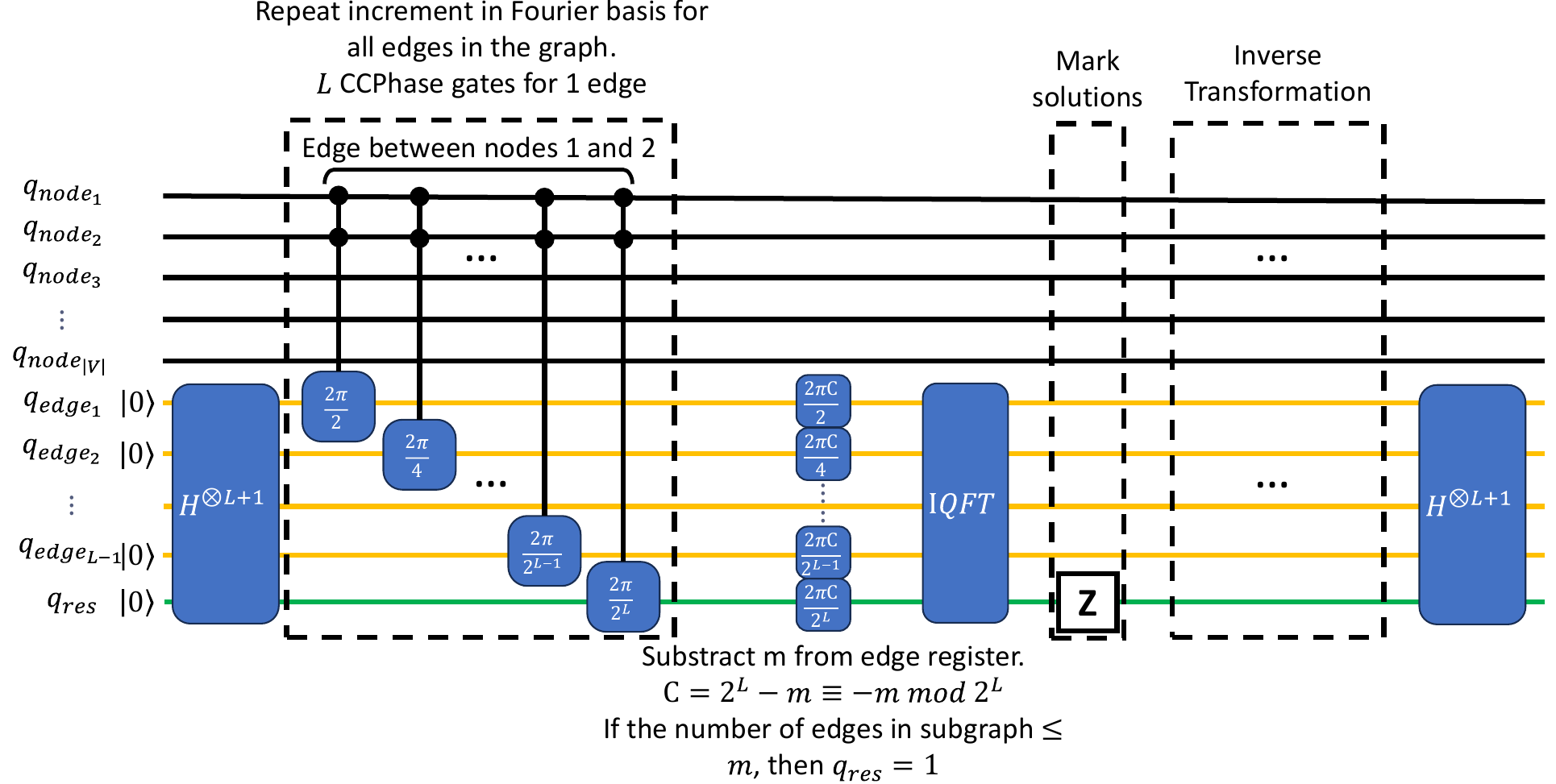}
    \caption{Gate scheme for oracle.}
    \label{fig:oracle}
\end{figure}

The oracle design for edge counting constitutes the most complex aspect of the quantum circuit, responsible for marking subgraphs containing at least the current threshold $m_i$ edges (see Fig. \ref{fig:oracle}).
The oracle implementation begins with allocation of an edge counter register ($q_{edge}$, $q_{res}$) consisting of $L + 1 = \lceil \log_2(\frac{k(k-1)}{2}) \rceil+1$ qubits, which provides sufficient capacity to store the maximum possible number of edges in any $k$-vertex subgraph. For each edge $(i,j)\in E$ in the original graph, the circuit applies controlled operations that increment the counter when both corresponding vertex qubits indicate inclusion in the subgraph. These controlled increment operations are optimized through implementation in the Quantum Fourier Transform basis, where increment operations translate to efficient phase rotations rather than complex arithmetic circuits.

The QFT-based counter implementation offers significant advantages in gate count and parallelism. The process begins with QFT preparation that converts the computational basis to the Fourier basis, achieved through Hadamard gates since the edge counting register is initialized in the $|0\rangle^{\otimes L+1}$ state. Phase accumulation follows, where for each potential edge the circuit applies controlled phase rotations to the counter qubits. After processing all edges, an inverse QFT returns the register to the computational basis for subsequent operations. The complete edge counting procedure requires $O(|E|L)$ gates, where $|E|$ represents the number of edges in the original graph.

Following edge counting, a quantum comparator circuit evaluates whether the counted edges meet or exceed the threshold $m_i$. This comparison employs two's complement arithmetic by computing $C = 2^{L+1} - m_i \equiv -m_i \mod 2^{L+1}$, then performing modular addition of $C$ to the edge count register. The most significant bit of the result indicates whether the original count exceeded $m_i$, providing the comparison outcome. The comparator circuit achieves high efficiency through $O(L)$ single-qubit gates applied simultaneously with constant unit depth.

The complete oracle operation ensures proper phase marking while maintaining reversibility through careful uncomputation. The process computes the edge count into the counter register, applies the comparator to flip the phase of states satisfying the edge threshold condition, then uncomputes both the comparator and edge counting operations. This sequence ensures the oracle leaves no residual entanglement except for the crucial phase marking of valid solutions.

\begin{figure}
    \centering
    \includegraphics[width=1\linewidth]{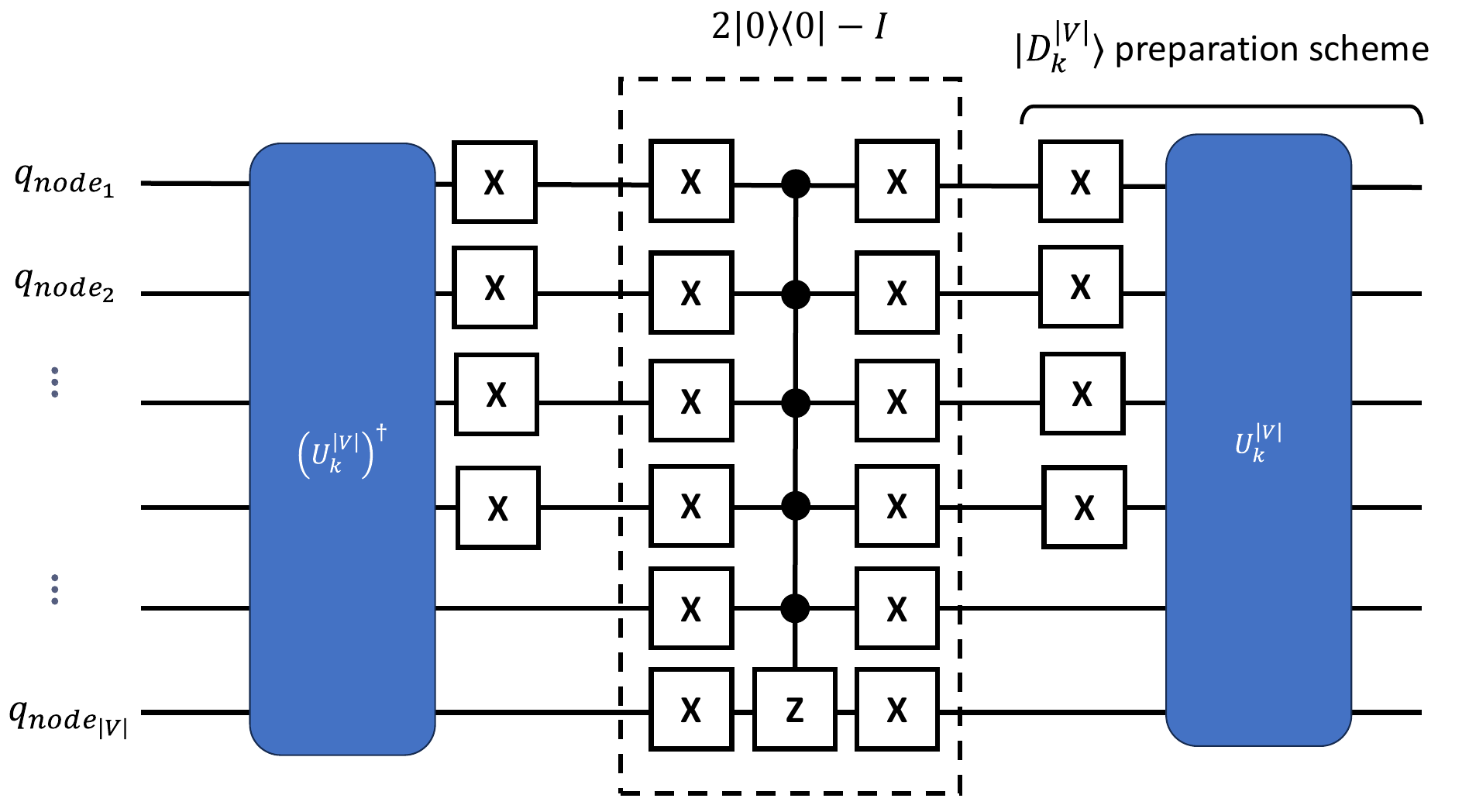}
    \caption{Diffusion operator}
    \label{fig:diffusion}
\end{figure}

The diffusion operator implementation (see Fig. \ref{fig:diffusion}) completes the Grover iteration by reflecting the state about the initial Dicke state according to $U_{\text{diff}} = 2|D^n_k\rangle\langle D^n_k| - I$. 
This operator is constructed as $U_{\text{diff}} = U^n_k X^{\otimes n} (2|0^n\rangle\langle 0^n| - I) X^{\otimes n} (U^n_k)^\dagger$, where $U^n_k$ represents the Dicke state preparation circuit and $X^{\otimes n}$ flips all qubits. The core reflection operation $2|0^n\rangle\langle 0^n| - I$ is implemented using an $n$-controlled $Z$ gate with $O(n)$ overhead through careful use of ancillary qubits. The complete diffusion operator requires $O(nk)$ gates, dominated by the Dicke state preparation component.

The resource requirements for the complete quantum circuit are determined by several factors. The vertex register requires $n$ qubits, one for each vertex in the original graph. The edge counter demands $L + 1 = \lceil \log_2(\tfrac{k(k-1)}{2}) \rceil$ qubits to accommodate the maximum possible edge count.
The gate complexity per Grover iteration scales as $O(|E|L + kn)$, with the dominant contributions coming from edge counting and Dicke-state preparation; the corresponding circuit depth is $O(|E| + k \log (n/k))$ for all-to-all connectivity.

To handle the unknown number of solutions at each threshold level, we employ the Dürr–Høyer modification of Grover's algorithm~\cite{durr1996quantum}, which guarantees a success probability of at least $1/4$ per search attempt. This probabilistic guarantee arises from the uniform random selection of iteration counts $t$ from the interval $[0, \frac{\pi}{4}\sqrt{N}]$, where $N = \binom{n}{k}$ is the size of the search space. For a fixed number $M$ of marked $k$-subsets, the success probability after $t$ Grover iterations is $P(t) = \sin^2((2t+1)\theta)$ with $\sin^2\theta = M/N$, and averaging over the random choice of $t$ yields
\begin{equation}
\frac{1}{T}\sum_{t=0}^{T-1} \sin^2((2t+1)\theta) \ge \frac{1}{4},
\end{equation}
for $T = \left\lceil \frac{\pi}{4}\sqrt{N}\right\rceil$ and all $\theta$. This ensures that after $R$ consecutive failures we can terminate with high confidence that no better solution exists, following the criterion $(1-s)^R \le 1-p$ with $s \ge 0.25$ and $p$ representing our desired confidence level. Solving this inequality yields $R \ge \frac{\log(1 - p)}{\log(1 - s)} \approx 11$ iterations for $p=95\%$.

Combining this with the cost of a single Grover iteration gives an overall bound on the number of oracle calls. For a fixed threshold value $m_i$, each Dürr–Høyer attempt uses on average $T/2 = O(\sqrt{N})$ Grover iterations, and hence $O(\sqrt{N})$ oracle calls. Since the success probability of an attempt is bounded below by $s \ge 1/4$, a constant number $R = O(1)$ of independent attempts suffices to reach a prescribed confidence that either a denser $k$-subgraph has been found or none exists at this threshold. Thus, the expected oracle cost per threshold level is $O(\sqrt{N})$. Searching over all threshold values $m_i$ requires at most
$K \le k(k-1)/2$ distinct levels, because the thresholds are updated
to strictly larger edge counts and therefore form an increasing
sequence of integer values in the range $[m_0, k(k-1)/2]$. In principle, a binary search over the possible edge counts can reduce this to $K = O(\log k)$, but in both cases the dominant contribution to the complexity comes from the $\sqrt{N} = \sqrt{\binom{n}{k}}$ factor arising from Grover-type amplitude amplification. Consequently, the total number of oracle calls needed to solve the densest-$k$-subgraph problem on a graph $G=(V,E)$ with $|V|=n$ satisfies
\begin{equation}
O\bigl(K\sqrt{N}\bigr) = O\!\left(k^2 \sqrt{\binom{n}{k}}\right)
\end{equation}
for threshold-by-threshold search, or $O\!\bigl(\log k \sqrt{\binom{n}{k}}\bigr)$ when using binary search over $m_i$.

In particular, for fixed $k$ this yields a quadratic speedup over Brute-force, which in the worst case must evaluate all $\binom{n}{k}$ candidate $k$-subgraphs.

This comprehensive implementation details all quantum circuit components necessary for solving the densest $k$-subgraph problem, providing both the theoretical foundation and a concrete circuit-level framework for its implementation.

\section{Numerical Simulations}
We evaluate the Grover-based search in two complementary settings. First, we study convergence trajectories on a fixed small graph. Second, we perform a scaling study on randomly generated graphs with varying $n$ and $k$, measuring how the oracle budget required to reach the optimum grows with the search-space size $N=\binom{n}{k}$.

We compare four search procedures for the densest-$k$-subgraph problem: \emph{Quantum Grover}, \emph{Black-box Grover emulator}, \emph{Brute-force}, and \emph{Simulated annealing}. 
Quantum Grover denotes a noiseless quantum simulation of the algorithm using the Qiskit \texttt{AerSimulator} backend, implementing the full Grover iteration schedule described above. 
The Black-box Grover emulator uses the same adaptive optimization procedure, but instead of explicitly simulating the quantum circuit, the Grover search subroutine is replaced by a classical black-box routine. Whenever a marked $k$-subset exists, this routine returns the first such subset found by Brute-force search with probability exactly $25\%$, and with the remaining probability returns a uniformly random $k$-subset. This probability matches the theoretical worst-case lower bound on the success probability of our Grover-based procedure.
Brute-force corresponds to classical exhaustive search over all $\binom{n}{k}$ subsets and provides the ground-truth optimum used as a reference. 
Finally, Simulated annealing is a classical heuristic baseline with a $k$-swap neighborhood and a tabu-search mechanism, which improves its ability to escape local minima; each run is initialized from a uniformly random $k$-vertex subset.

To enable a fair comparison, we express computational effort in units of \emph{oracle calls}. One oracle call is defined as a single evaluation of the objective function, i.e., the edge count of a candidate $k$-vertex subset. For Quantum Grover, this corresponds to one application of the phase oracle within the Grover iteration schedule. The Black-box Grover emulator is counted in the same way: one oracle call per logical oracle query in the adaptive schedule, irrespective of how the corresponding marked subset is found internally. For the classical baselines (Brute-force and simulated annealing), one oracle call simply equals one evaluation of the edge count for a visited $k$-subset. Convergence is reported in terms of the best-so-far edge count observed after a given number of oracle calls.

\begin{figure}[t]
    \centering
    \includegraphics[width=\linewidth]{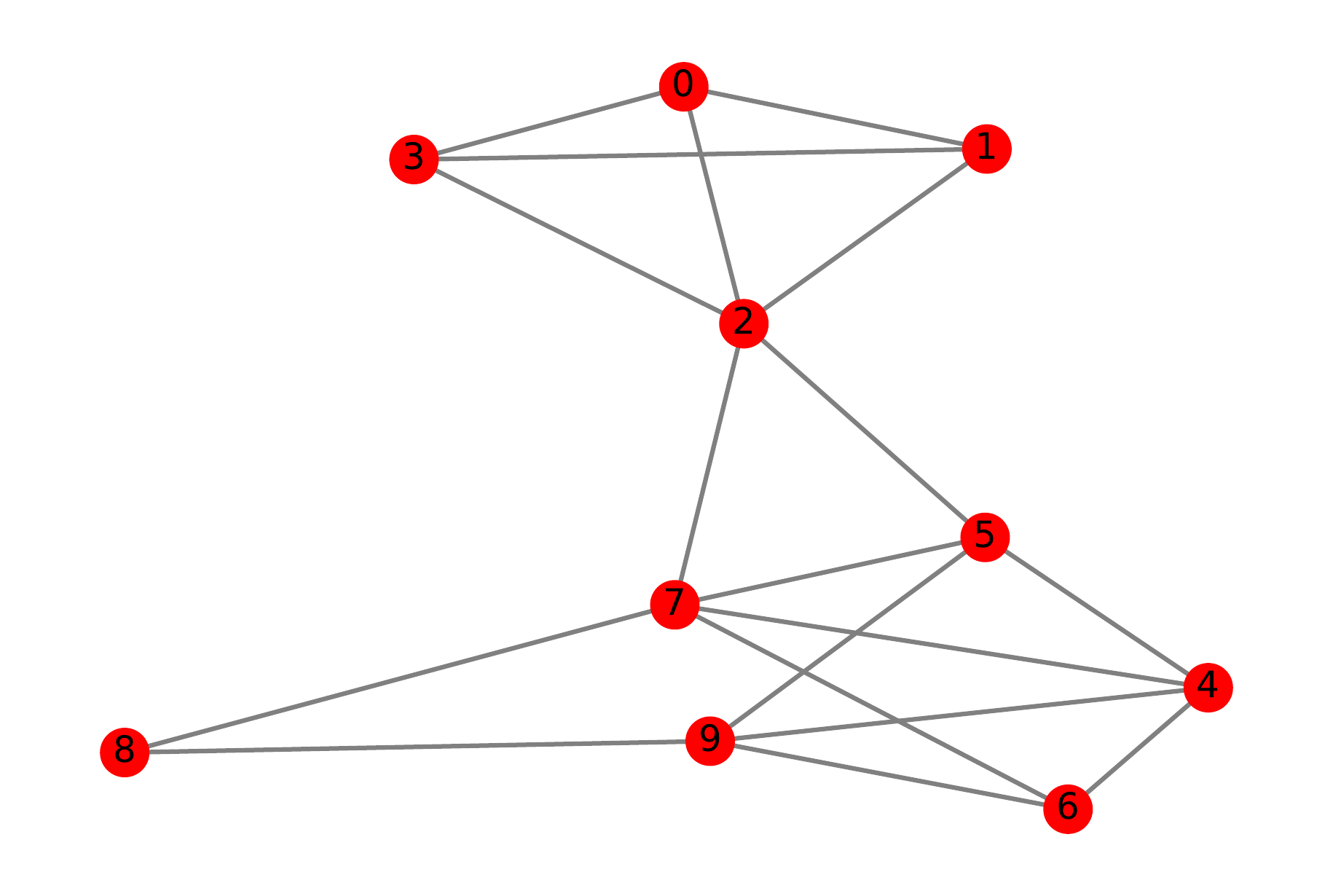}
    \caption{Benchmark graph with $n = 10$ vertices used in the convergence experiments.}
    \label{fig:sample_graph}
\end{figure}

\begin{figure*}[!t]
    \centering
    \includegraphics[width=0.32\textwidth]{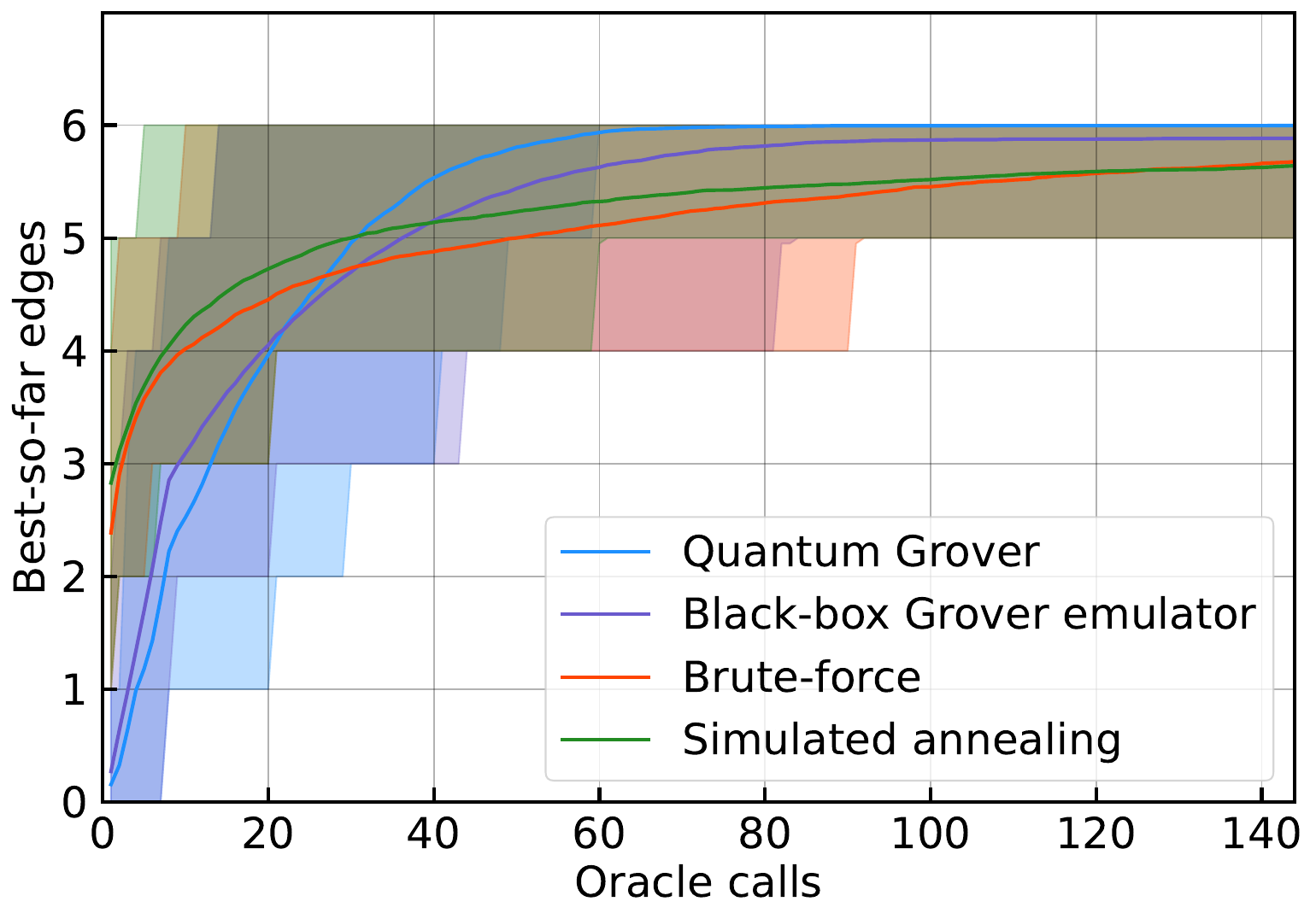}
    \includegraphics[width=0.32\textwidth]{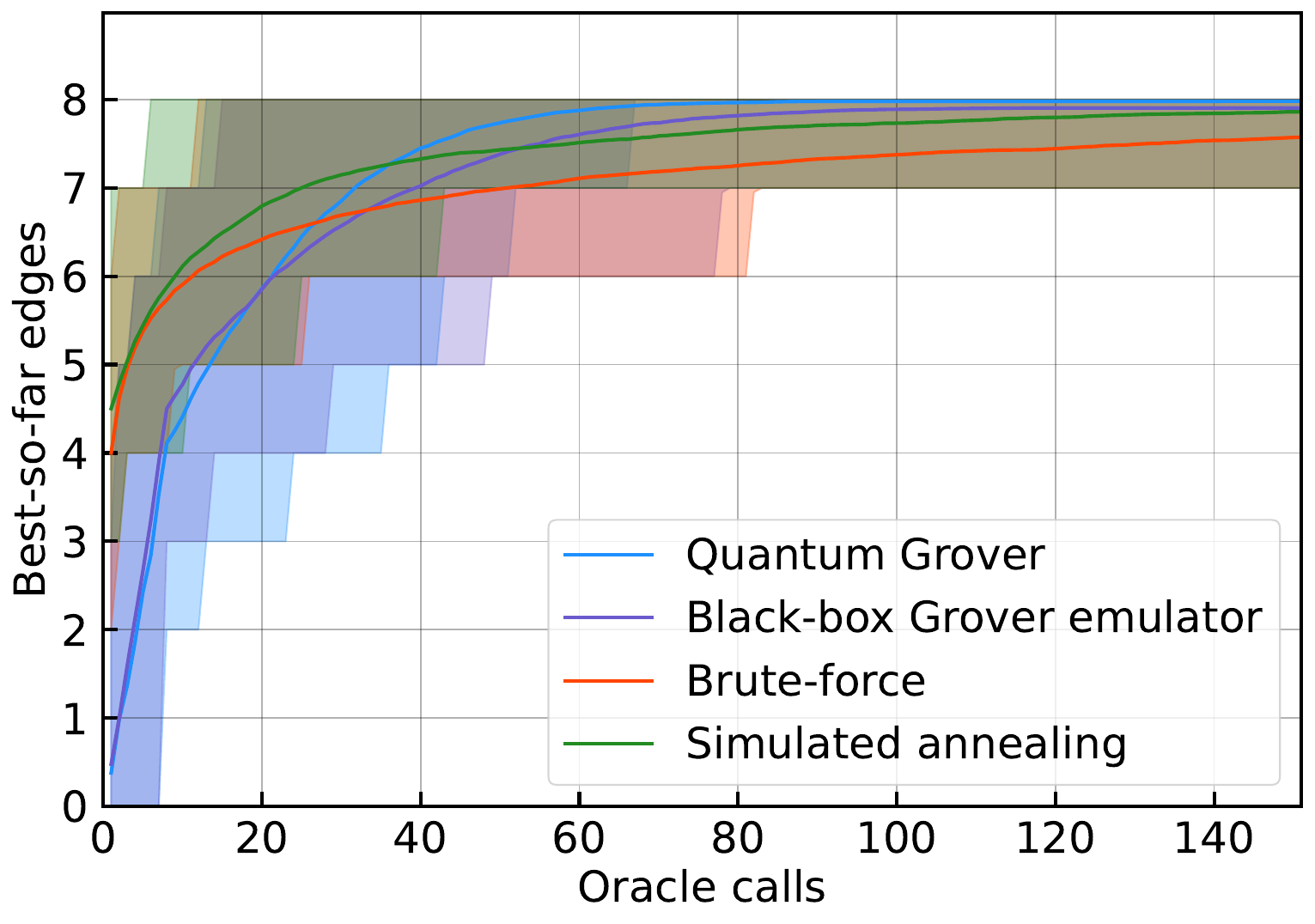}
    \includegraphics[width=0.32\textwidth]{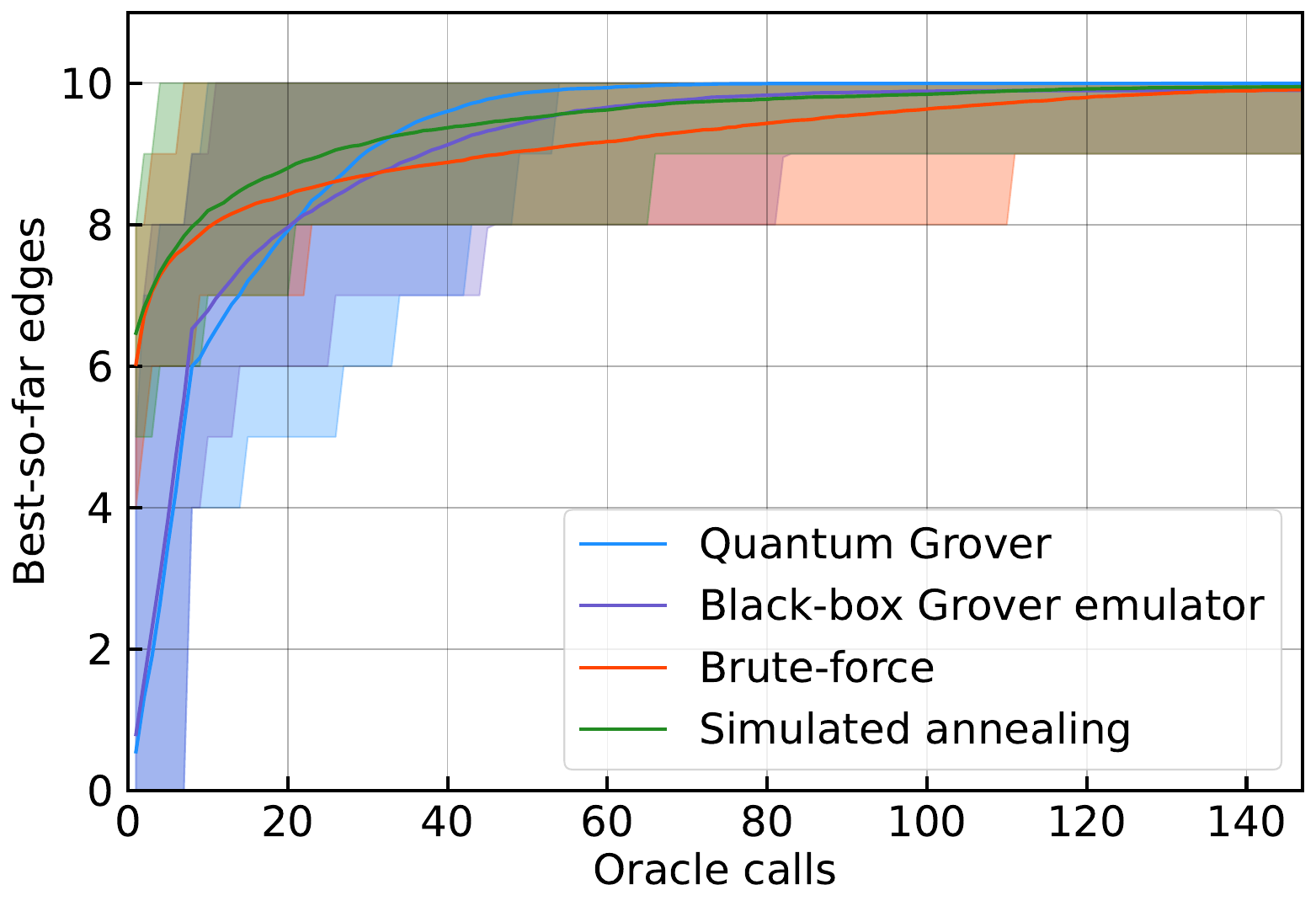}
    \caption{Convergence of the best-so-far edge count for the four algorithms on the benchmark graph from Fig.~\ref{fig:sample_graph}. From left to right: $k = 4,5,6$. Solid lines show the mean over 1000 independent runs, and shaded regions indicate empirical $90\%$ confidence intervals.}
    \label{fig:search_examples}
\end{figure*}

As a benchmark instance, we consider a fixed undirected graph on $n = 10$ vertices. We use this graph to compare the convergence of all four algorithms for subgraph sizes $k \in \{4,5,6\}$; the instance is shown in Fig.~\ref{fig:sample_graph}.
For this benchmark graph, we run each algorithm for $k \in \{4,5,6\}$ and, at every oracle call, record the best-so-far edge count. Figure~\ref{fig:search_examples} shows the resulting convergence curves as a function of the number of oracle calls: each panel corresponds to a different value of $k$, solid lines denote the mean over 1000 independent runs, and shaded bands represent empirical $90\%$ confidence intervals.

\begin{figure}[t]
    \centering
    \includegraphics[width=\linewidth]{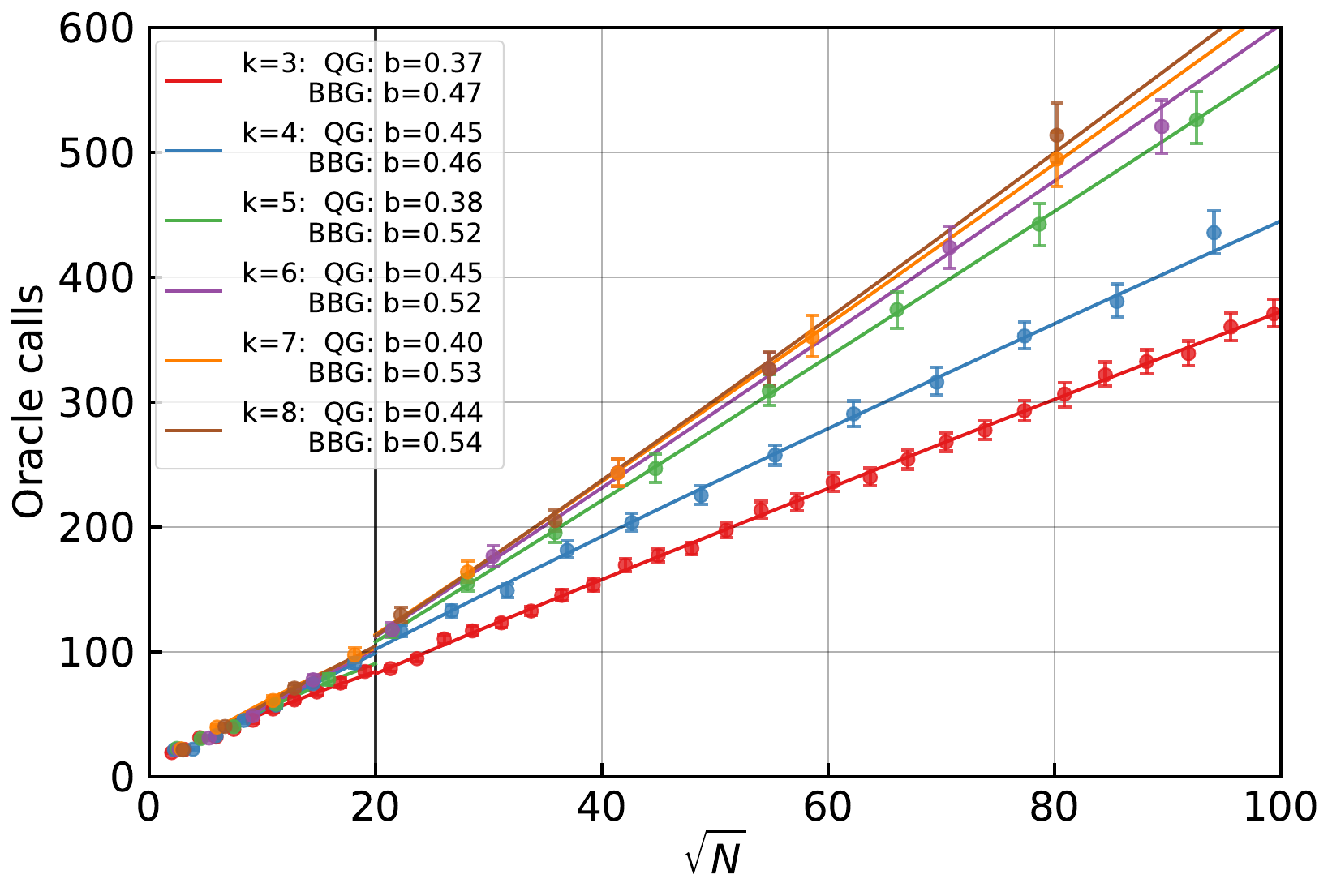}
    \caption{Scaling of the oracle cost of the Grover-based search as a function of the search-space size $N = \binom{n}{k}$ for several values of $k$. Each point shows the average number of oracle calls needed to certify the optimal densest $k$-subgraph with probability at least $95\%$; error bars indicate $99\%$ hierarchical bootstrap confidence intervals. Here QG denotes the full Quantum Grover simulation and BBG the Black-box Grover emulator. The lines show power-law fits of the form $y = aN^b$, with the fitted exponent $b$ given in the legend.}
    \label{fig:results}
\end{figure}

To study the scaling behaviour, we measure how many oracle calls are required to find the densest $k$-subgraph as a function of the search-space size $N = \binom{n}{k}$, i.e., the total number of $k$-vertex subsets. For instances with $\sqrt{N} \le 20$ we use the Quantum Grover simulation, while for larger instances ($\sqrt{N} > 20$) we switch to the Black-box Grover emulator, which allows us to reach significantly larger values of $N$ at the oracle level. For each $k \in \{3,4,5,6,7,8\}$ and each admissible $n$, we generate $20$ independent Erdős--Rényi \cite{renyi1959random} graphs $G(n,0.5)$ and, for every graph, perform $20$ runs of the search algorithm. The Grover-based search is terminated when the estimated probability that a denser $k$-subgraph still exists drops below $5\%$ (equivalently, when the current best candidate is optimal with probability at least $95\%$). The resulting scaling of the oracle cost with $N$ is shown in Fig.~\ref{fig:results}. In that figure, QG denotes the full Quantum Grover simulation and BBG the Black-box Grover emulator. The curves are fitted by power laws of the form $y = aN^b$, and the fitted exponent $b$ is reported in the legend.

To quantify statistical uncertainty, we apply a two-level bootstrap with $n_{\text{boot}} = 2000$ resamples. This accounts for the hierarchical structure of the data: repeated runs are nested within each graph, and multiple graphs are generated for each pair $(n,k)$. For each bootstrap sample, we first resample runs within each fixed graph and aggregate them using the median number of oracle calls, and then resample graphs and average the resulting per-graph medians. This yields a bootstrap distribution of the oracle cost for every pair $(n,k)$, from which we extract the mean and a $99\%$ confidence interval.

\begin{figure}[t]
    \centering
    \includegraphics[width=\linewidth]{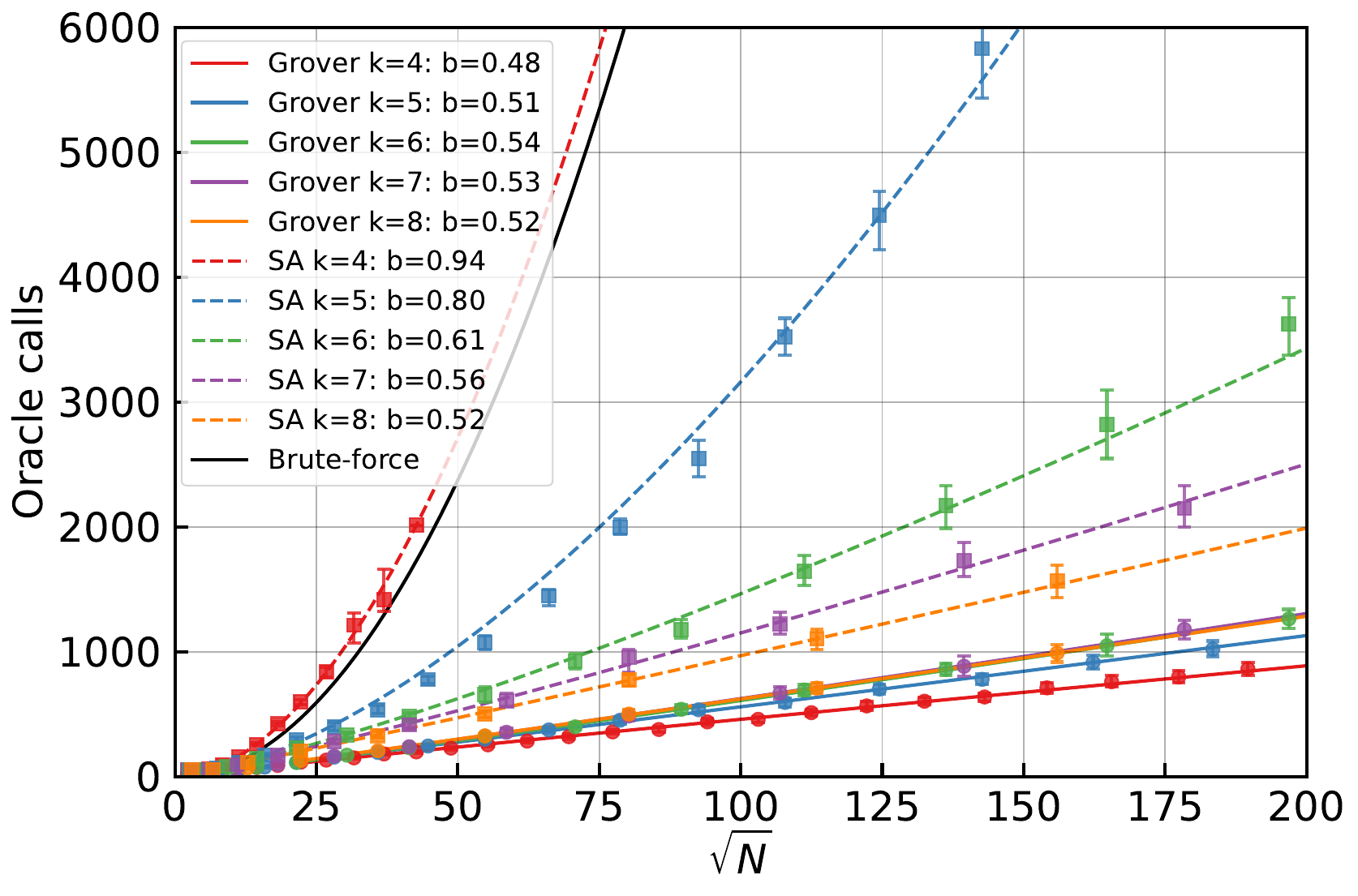}
    \caption{Comparison of oracle cost scaling for the Grover-based search and classical baselines as a function of the search-space size $N = \binom{n}{k}$. Circles show the Grover-based oracle cost needed to reach a $95\%$ success probability, while square markers correspond to Simulated annealing tuned to the same target success probability. The black curve represents classical Brute-force search with $95\%$ success probability. The colored lines show power-law fits of the form $y = aN^b$, with the fitted exponent $b$ given in the legend.}
    \label{fig:results_classical}
\end{figure}

To compare this quantum scaling with classical baselines, we also estimate the oracle cost of Brute-force search and Simulated annealing (SA) as a function of $N$, as shown in Fig.~\ref{fig:results_classical}. For Brute-force, one oracle call corresponds to evaluating the edge count of a single $k$-subset, and to reach a $95\%$ success probability, we take the expected oracle cost to be $0.95\,N$, assuming that subsets are examined in random order. For SA, for each pair $(n,k)$, we generate $100$ Erdős--Rényi graphs of density $0.5$ conditioned to have a unique densest $k$-vertex subgraph. For every such graph, we run the algorithm $1000$ times to estimate the single-run success probability $s$ of finding this subgraph. The number of independent SA runs required to achieve an overall success probability of at least $95\%$ is then
\begin{equation}
T = \max\!\left(1, \frac{\log(1-0.95)}{\log(1-s)}\right),
\end{equation}
so that the corresponding oracle cost is $\lceil T d \rceil$, where $d$ is the number of oracle calls per SA run. Averaging these costs over graphs yields the SA data points. The Grover and SA datasets are again fitted by power laws of the form $y = aN^b$, with the fitted exponents shown in the legend. The fitted exponents are consistent with the expected near-$\sqrt{N}$ scaling of the Grover-based search, while Brute-force grows linearly in $N$; the SA baseline exhibits intermediate behaviour over the tested range.

\section{Discussion}
The results presented in this work show that the DkS problem can be approached from two distinct algorithmic perspectives. A standard formulation is the QUBO representation, which has become a common framework in quantum approaches to NP-hard combinatorial optimization problems\cite{calude2020quantum,denchev2016computational} and is reflected here in the simulated annealing baseline. In contrast, the Grover-based construction developed in this work treats DkS directly as a constrained search problem over $k$-vertex subsets, without relying on a QUBO encoding as the primary algorithmic framework. In this sense, our approach is closer to the broader line of research exploring non-QUBO quantum algorithms for graph problems\cite{ambainis2019quantum,applied_quantum_walks2021}.


From a conceptual standpoint, the Grover-based algorithm provides an explicit search mechanism with well-defined performance guarantees. Unlike quantum annealing, which relies on penalty-based energy landscapes and is sensitive to embedding choices and analog noise\cite{choi2008minor,king2021scaling}, our method operates on a well-defined Hilbert subspace consisting exactly of $k$-vertex subsets prepared via Dicke states\cite{bartschi2020dicke}. This eliminates heuristic hyperparameters such as penalty weights, whose improper tuning can obscure optimal solutions or bias the search process. The explicit circuit structure of our oracle, based on QFT-based edge counting\cite{draper2000addition}, makes the structure of the computation transparent and analyzable.

However, these advantages come at the cost of deeper circuits and more demanding resource requirements. The oracle design incurs substantial overhead, especially for dense input graphs, where the number of graph edges $|E|$ becomes large. Each graph edge contributes a sequence of controlled-phase rotations, whose T-count dominates the logical resource requirements in fault-tolerant settings\cite{gidney2021magic}. Consequently, while the Grover approach achieves asymptotic quadratic speedup over classical brute force, this advantage is accompanied by large constant-factor overheads.

Despite these limitations, the potential for improvement is significant. Approximate Dicke state preparation techniques\cite{zoufal2022variational} could substantially reduce the circuit depth while still maintaining a nearly uniform distribution over $k$-subsets. Alternative edge-counting mechanisms, such as carry-save arithmetic\cite{thapliyal2013reversible} or blockwise parallel aggregation inspired by recent quantum adder optimizations\cite{haner2018optimizing}, may reduce the dependence on QFT-based counters. Similarly, more advanced Grover iteration schedules---e.g., based on amplitude estimation\cite{brassard2002quantum} or quantum minimum finding heuristics\cite{naderipour2023improving}---could reduce the number of required Grover iterations.

Classical--quantum hybrid strategies offer another promising direction. Classical preprocessing can substantially reduce the size of the search space by pruning low-degree vertices, identifying locally dense regions of the graph, or generating high-quality warm starts, as observed in hybrid quantum algorithms for related combinatorial tasks\cite{egger2021warmstart,harrigan2021quantum}. Such a hybrid strategy aligns naturally with the DkS problem, where classical heuristics can often identify promising regions of the search space before the quantum stage.

Looking ahead, the transition to fault-tolerant hardware will be an important factor in assessing the practicality of different quantum approaches to DkS. While quantum annealing currently appears more scalable on near-term devices, gate-based approaches benefit from rapid algorithmic development and natural compatibility with error-corrected logical operations. Continued improvements in circuit synthesis, qubit connectivity, and error correction may further improve the prospects of algorithms such as the one proposed here. Overall, the existence of multiple quantum approaches broadens the range of tools that can be matched to the structure of a given problem.

\section{Conclusion}

In this work, we have developed and analyzed an explicit quantum search algorithm for the Densest~$k$-Subgraph (DkS) problem, demonstrating that quantum techniques beyond the conventional QUBO formulation can be effectively applied to this central combinatorial optimization task. By constructing all components of the algorithm---including deterministic Dicke-state preparation, a fully reversible QFT-based edge-counting oracle, and a diffusion operator tailored to the constrained search space---we provide a complete gate-level description suitable for both theoretical analysis and prospective implementation on future quantum hardware.

Our approach achieves a provable quadratic speedup over classical Brute-force search through amplitude amplification. Unlike QUBO-based quantum annealing methods, it avoids penalty parameters and provides explicit control over the search subspace, thereby ensuring transparent resource estimation and predictable behavior. Numerical simulations further illustrate the algorithm's convergence characteristics and confirm that, for moderate system sizes, the Grover framework begins to outperform classical Brute-force methods in terms of oracle queries.

Beyond the concrete algorithm presented here, this work demonstrates that explicit quantum circuit constructions constitute a viable and powerful alternative to QUBO embeddings for NP-hard problems. The techniques developed---Dicke-state initialization, reversible edge counting, and structured reflections---are modular and readily applicable to related graph-theoretic problems such as maximum clique or dense community detection. These building blocks therefore help broaden the palette of quantum primitives available for optimization on near-term and fault-tolerant hardware.

While the resource requirements of the current construction exceed the capabilities of contemporary NISQ devices, the methods outlined in this work provide a clear roadmap for future investigations. Possible avenues for improvement include approximate or variational Dicke-state preparation, optimized arithmetic circuits for edge counting, hybrid classical--quantum preprocessing strategies, and refined thresholding schemes in the spirit of amplitude estimation. As fault-tolerant quantum processors mature, algorithms like the one developed here may play an important role in realizing practical quantum advantage for complex network-analysis tasks.

\section{Acknowledgment}
The work was supported by Russian Science Foundation grant 22-12-00353-$\Pi$ (https://rscf.ru/en/project/22-12-00353/). The work was also supported by Rosatom in the framework of the Roadmap for Quantum computing (Contract No. 868-1.3-15/15-2021 dated October 5, 2021). Yu.A. B is grateful to the Russian Foundation for the Advancement of Theoretical Physics and Mathematics (BASIS) (Projects №24-2-10-57-1).

\bibliography{references}
\end{document}